\documentstyle [12pt] {article}
\title {A SIMPLE SOLUTION OF THE  LOTKA-VOLTERRA EQUATIONS }

\author {Vladan Pankovi\'c $^{\ast,\sharp}$, Dejan Banjac$^{\sharp}$, Rade Glavatovi\'c $^{\Diamond}$,Milan Predojevi\'c $^{\ast,\sharp}$\\
$^{\ast}$ Department of Physics, Faculty of Sciences \\21000 Novi Sad, Trg Dositeja Obradovi\'ca 4,  Serbia \\
$^{\sharp}$ Gimnazija, 22320 Indjija, Trg Slobode 2a \\ Serbia, vladanp@gimnazija-indjija.edu.yu\\
$^{\Diamond}$Military Medical Academy, 11000 Beograd, Crnotravska
17, Serbia\\}

\date {}
\begin {document}
\maketitle

\vspace {1cm}

\begin {abstract}
In this work we consider a simple, approximate, tending toward
exact, solution of the system of two usual Lotka-Volterra
differential equations. Given solution is obtained by an iterative
method. In any finite approximation order of this solution,
exponents of the corresponding Lotka-Volterra variables have
simple, time polynomial form. When approximation order tends to
infinity obtained approximate solution  converges toward exact
solution in some finite time interval.
\end {abstract}
\vspace {1.5cm}

\section {Introduction}

As it is well-known [1]-[5] the following system of two nonlinear
differential equations of the first order
\begin {equation}
   \frac {dx}{dt} = ax - bxy
\end {equation}
\begin {equation}
   \frac {dy}{dt} = -cy + dxy
\end {equation}
represents the remarkable, usual Lotka-Volterra  differential
equations system with significant applications in the ecology,
biology, medicine etc.. Here $x$, $y$, represent the real,
positive variables that depends of the time, $t$, while $a$, $b$,
$c$, $d$, represent real, positive time independent constants,
i.e. parameters. It is supposed that the initial values
\begin {equation}
   x(0) = x_{0}
\end {equation}
\begin {equation}
   y(0) = y_{0}
\end {equation}
of $x$, $y$ are known. Existing, exact solution of the system (1),
(2) presentable in the simple (closed) form (combination of the
simple functions of the time, eg. a polynomial of the time etc.)
is unknown to this day. For this reason different approximate,
especially numerical (Runge-Kutta etc.), methods for solution of
the system (1), (2) are used. But, even in this case approximate
solution of the system (1), (2) presentable in a simple form is
unknown.

In this work we shall suggest a simple, approximate, tending
toward exact, solution of the Lotka-Volterra equations system (1),
(2). Given solution will be obtained by an iterative method. In
any finite approximation order of this solution, exponents of the
corresponding Lotka-Volterra variables have simple, time
polynomial form. When approximation order tends to infinity
obtained approximate solution  converges toward exact, simple
solution in some finite time interval. All this can be very useful
for the applications in many domains of the ecology, biology,
medicine, etc., but explicit consideration of such applications
goes over basic intentions of this work.

\section { A simple approximate solution of the usual Lotka-Volterra equations system }

As it is well-known [1]-[5] usual Lotka-Volterra equations system
(1), (2) can be simply transformed in the following system
\begin {equation}
  \frac {d \ln x }{dt} = a - by
\end {equation}
\begin {equation}
   \frac {d \ln y }{dt} = -c + dx
\end {equation}
or, after well-known changes of the variables,
\begin {equation}
    u = \ln x      \Leftrightarrow    x = \exp u
\end {equation}
\begin {equation}
    v = \ln y      \Leftrightarrow    y = \exp v
\end {equation}
(so that $u$, $v$ represent the real variables, exponents of $x$,
$y$), in the system
\begin {equation}
     \frac {du}{dt} = a - b\exp v
\end {equation}
\begin {equation}
   \frac {dv}{dt} = -c + d\exp u
\end {equation}
It is supposed that the initial values
\begin {equation}
  u(0) = u_{0}
\end {equation}
\begin {equation}
  v(0) = v_{0}
\end {equation}
of $u$, $v$ corresponding to initial values $x_{0}$, $y_{0}$ (3),
(4) are known.

Solution of   the system (9), (10) for conditions (11), (12) will
be supposed in the following approximate form  characteristic for
$n$-th approximation step
\begin {equation}
    u_{(n)} = A_{0} + A_{1}t + A_{2}t^{2}+ . . . + A_{n}t^{n}    \hspace{0.5cm} {\rm for} \hspace{0.5cm} n = 0, 1, 2, …
\end {equation}
\begin {equation}
    v_{(n)} = B_{0}+ B_{1}t + B_{2}t^{2} + . . . + B_{n}t^{n}      \hspace{0.5cm}  {\rm for} \hspace{0.5cm} n = 0, 1, 2, …
\end {equation}
where $A_{n}$ and $B_{n}$ for $n = 0,1,2$. . . . are unknown real
coefficients. These coefficients will be determined by supposition
\begin {equation}
    A_{0} = u_{0}
\end {equation}
\begin {equation}
    B_{0} = v_{0}
\end {equation}
and by approximate iterative method
\begin {equation}
\frac {du_{(n)}}{dt} = a - b \exp [v_{(n-1)}]  \hspace{0.5cm} {\rm
for} \hspace{0.5cm} n =  1,2,. . . . .
\end {equation}
\begin {equation}
\frac {dv_{(n)}}{dt} = -c + d \exp [u_{(n-1)}] \hspace{0.5cm} {\rm
for} \hspace{0.5cm} n =  1,2,. . . . .
\end {equation}

In the first approximation order, i.e. for $n = 1$, introduction
of (15),(16) in (17),(18) yields immediately
\begin {equation}
    A_{1} = a - b \exp[v_{0}]
\end {equation}
\begin {equation}
    B_{1} = -c + d exp[u_{0}]
\end {equation}
In this way coefficients $A_{1}$ and $B_{1}$ are completely
determined.

In the second approximation order, i.e. for $n = 2$, introduction
of (15),(16),(19),(20) in (17),(16) yields
\begin {equation}
  A_{1} + 2A_{2}t = a - b \exp [v_{0} + B_{1}t] = a - b \exp [v_{0}] \exp [B_{1}t]
\end {equation}
\begin {equation}
  B_{1} + 2B_{2}t = -c + d \exp [u_{0} + A_{1}t]  =  - c + d \exp [u_{0} ] \exp [A_{1}t]
\end {equation}
For additional conditions
\begin {equation}
    | B_{1}t | \ll 1
\end {equation}
\begin {equation}
    | A_{1}t | \ll 1
\end {equation}
i.e. for
\begin {equation}
    t \ll |\frac {1}{B_{1}}|
\end {equation}
\begin {equation}
    t \ll |\frac {1}{A_{1}}|
\end {equation}
right hands of (21), (22) can be approximated by theirs linear
Taylor expansions which yields
\begin {equation}
  A_{1} + 2A_{2}t = a - b \exp [v_{0}](1+ B_{1}t)
\end {equation}
\begin {equation}
  B_{1} + 2B_{2}t =  -c+d \exp [u_{0}](1+A_{1}t)
\end {equation}
It, according to (19), (20), yields
\begin {equation}
   A_{2} = - \frac {b B_{1} \exp [v_{0}]}{2}
\end {equation}
\begin {equation}
   B_{2} =  \frac {d A_{1} \exp [u_{0}]}{2}
\end {equation}
In this way coefficients $A_{2}$ and $B_{2}$ are completely
determined.

In $n+1$-th approximation order introduction of (13), (14) in
(17), (18) yields
\begin {eqnarray}
  A_{1} + 2A_{2}t +. . . + (n+1)A_{n+1}t^{n} = a - b \exp [v_{0} + B_{1}t + B_{2}t^{2} + . . . + B_{n}t^{n} ] = \nonumber \\
  =a - b \exp [v_{0} ] \exp [B_{1}t + B_{2}t^{2} + … +B_{n}t^{n}]  \hspace{0.5cm}  {\rm for} \hspace{0.5cm} n =  2,
  3,. . . .
\end {eqnarray}

\begin {eqnarray}
  B_{1} + 2B_{2}t  + . . .  + (n+1)B_{n+1}t^{n} = -c + d \exp [u_{0} + A_{1}t + A_{2}t^{2} + . . .  + A_{n}t^{n}]  = \nonumber \\
  = - c + d \exp [u_{0} ] \exp [A_{1}t + A_{2}t^{2} + … + A_{n}t^{n}] \hspace{0.5cm} {\rm for} \hspace{0.5cm} n =  2,
  3,. . .
\end {eqnarray}

Also, suppose that the following additional approximation
conditions are satisfied
\begin {equation}
 1 \gg | B_{1}t + B_{2}t^{2} + . . . + B_{n}t^{n}|   \hspace{0.5cm}  {\rm for} \hspace{0.5cm} n =  2,
 3,. . .
\end {equation}
\begin {equation}
 1 \gg | B_{1}t| \gg |B_{2}t^{2}| \gg  . . .  \gg |B_{n}t^{n}|  \hspace{0.5cm}  {\rm for} \hspace{0.5cm} n =  2,
 3,. . .
\end {equation}
\begin {equation}
 1 \gg | A_{1}t + A_{2}t^{2} + . . . + A_{n}t^{n}| \hspace{0.5cm}  {\rm for} \hspace{0.5cm}  n =  2,
 3, . . .
\end {equation}
\begin {equation}
 1 \gg | A_{1}t| \gg |A_{2}t^{2}| \gg . . . \gg |A_{n}t^{n}|  \hspace{0.5cm}  {\rm for} \hspace{0.5cm}  n =  2,
 3,. . .
\end {equation}
Obviously, condition  (34) ensures the convergence of (14), while
condition (36) ensures the convergence of the (13). As it is not
hard to see, all conditions (33)-(36) can be changed by one
condition
\begin {equation}
  t \ll min (|\frac {1}{B_{1}}|,|\frac {B_{1}}{B_{2}}|,. . ., |\frac {B_{n-1}}{B_{n}}|, |\frac {1}{A_{1}}|,|\frac {A_{1}}{A_{2}}|, . . ., |\frac {A_{n-1}}{A_{n}}|)  \hspace{0.5cm} {\rm for} \hspace{0.5cm}  n =  2, 3, …
\end {equation}
where right hand of the inequality (37) represents the minimum
function equivalent to the minimal arguments.

According to (33)-(36), right hands of  (31),(32) can be
approximated by theirs linear Taylor expansions which yields
\begin {equation}
  A_{1} + 2A_{2}t +. . .+ nA_{n}t^{n} = a - b \exp [v_{0} ] [1 + B_{1}t + B_{2}t^{2} +. . .+B_{n}t^{n}]   \hspace{0.5cm} {\rm for} \hspace{0.5cm} n =  2,
  3,. . .
\end {equation}
\begin {equation}
  B_{1} + 2B_{2}t  + . . .+ nB_{n}t^{n} = - c + d \exp [u_{0} ][1 +A_{1}t + A_{2}t^{2} +  . . .+ A_{n}t^{n}] \hspace{0.5cm} {\rm for} \hspace{0.5cm} n =  2,
  3,. . .
\end {equation}
Then, it follows
\begin {equation}
   A_{n} =  \frac {-b B_{n-1} \exp [v_{0}]}{n}  \hspace{0.5cm} {\rm for} \hspace{0.5cm} n = 2,
   3,. . .
\end {equation}
\begin {equation}
   B_{n} =  \frac {d A_{n-1} \exp [u_{0}]}{n}  \hspace{0.5cm} {\rm for} \hspace{0.5cm} n = 2,
   3,. . .
\end {equation}
or
\begin {equation}
   A_{n} = \frac {- bd \exp [u_{0}+v_{0}]}{n(n-1)}A_{n-2}  \hspace{0.5cm} {\rm for} \hspace{0.5cm} n = 2,
   3,. . .
\end {equation}
\begin {equation}
   B_{n} =  \frac {-bd  \exp [u_{0}+v_{0}]}{n(n-1)}B_{n-2}   \hspace{0.5cm} {\rm for} \hspace{0.5cm} n = 2,
   3,. . .
\end {equation}
Or,
\begin {equation}
   A_{2k} = \frac {(- bd \exp [u_{0}+v_{0}])^{k}}{(2k)!} A_{0}  \hspace{0.5cm} {\rm for} \hspace{0.5cm} k = 1, 2,
   3,. . .
\end {equation}
\begin {equation}
   A_{2k+1} = \frac {(- bd \exp [u_{0}+v_{0}])^{k}}{(2k+1)!} A_{1} \hspace{0.5cm}  {\rm for} \hspace{0.5cm} k = 1, 2,
   3,. . .
\end {equation}
\begin {equation}
   B_{2k} = \frac {(- bd \exp [u_{0}+v_{0}])^{k}}{(2k)!} B_{0}  \hspace{0.5cm}  {\rm for} \hspace{0.5cm} k = 1, 2,
   3,. . .
\end {equation}
\begin {equation}
   B_{2k+1} = \frac {(- bd \exp [u_{0}+v_{0}])^{k}}{(2k+1)!} B_{1}   \hspace{0.5cm}  {\rm for} \hspace{0.5cm} k = 1, 2,
   3,. . .
\end {equation}

In this way coefficients $A_{n}$ and $B_{n}$ are completely
determined by (44)-(47) for $n = 2, 3,$. . .  .

So, according to previous analysis, coefficients $A_{n}$ and
$B_{n}$ are completely determined for any $n$, i.e. for  $n =
0,1,2,$.... Obviously, coefficients $A_{n}$ and $B_{n}$ appear
explicitly in $n$-th approximation order but their form stand
conserved in any next approximation order, for $n = 0,1,2,$....

Further, from (40)-(47) it follows
\begin {equation}
   |\frac {A_{n}}{A_{n+1}}| ~ (n+1)   \hspace{0.5cm} {\rm for} \hspace{0.5cm} n = 2,
   3,. . .
\end {equation}
\begin {equation}
   |\frac {B_{n}}{B_{n+1}}| ~ (n+1) \hspace{0.5cm} {\rm for} \hspace{0.5cm} n = 2,
   3,. . .
\end {equation}
It admits that, in a rough approximation, (37) be reduced in
\begin {equation}
  t \ll min (|\frac {1}{B_{1}}|, |\frac {1}{A_{1}}|)
\end {equation}
It demonstrates roughly that suggested approximate solution (13),
(14) of  the system (9), (10) for conditions (11), (12) is
consistent and convergent in a  time interval  $[0, \tau]$ (whose
explicit form will not be considered here) that is not infinite
small, i.e. infinitesimal.

Obviously, when n tends toward infinity consistent and convergent
approximate solution (13), (14) of  the system (9), (10) for
conditions (11), (12) in a finite time interval $[0, \tau]$ tends
toward exact solution.

Finally, without detailed analysis, we can observe the following.
We do not consider here the explicit form of $[0, \tau]$, i.e.
$\tau$ in general case. Also,  explicit form of the period, $T$,
of the Lotka-Volterra periodical variables $x$, $y$ is unknown in
general case. For this reason we cannot compare directly $\tau$
and $T$ in general case. Nevertheless, we can generally suppose
either $\tau > T$ or $\tau \leq T$. In the first case suggested
solution (7), (8), (13)-(16), (19), (20), (44)-(47) represents a
complete solution of the Lotka-Volterra equations system (1), (2)
with initial conditions (3),(4). In the second case, according to
(7), (8), (13)-(16), (19), (20), (44)-(47), we can determine
exactly $x(\tau)$ and $y(\tau)$. Then, $x(t)$, $y(t)$ solutions of
(1), (2) in a time moment $t = \tau + t'$  for initial conditions
(3), (4), can be presented by  $x(t')$, $y(t')$ solution of (1),
(2) where $t'$ represents new (translated) time variable for new
initial conditions $x(t' = 0) = x(\tau)$ and $y(t'= 0) = y(\tau)$.
Given solution can be obtained by expressions analogous to (7),
(8), (13)-(16), (19),( 20), (44)-(47) and this solution is
consistent and convergent in some  new, finite time interval $[0,
\tau ']$. If $\tau +\tau '$ is larger than $T$ or equivalent to
$T$ obtained solution represents final solution. In opposite case
described procedure of the solution by "translation" of the time
variable must be further repeated. It can be supposed that final
solution would be obtained after finite number of the repetitions
of the suggested solution procedure. Simply speaking, exact,
simple solution of the usual Lotka-Volterra equations would be
obtained by a finite number of the suggested analytical
continuations.

\section { Conclusion }

In conclusion we can shortly repeat and point out the following.
In this work we  suggest a simple, approximate, tending toward
exact, solution of the usual Lotka-Volterra differential equations
system. Given solution is obtained by an iterative method. In any
finite approximation order of this solution, exponents of the
corresponding Lotka-Volterra variables have simple, time
polynomial form. When approximation order tends to infinity
obtained approximate solution  converges toward exact solution in
some finite time interval. All this can be very useful for the
applications in many domains of the ecology, biology, medicine,
etc., but such applications are not considered explicitly in this
work.

\section { References }

\begin {itemize}
\item [ [1] ]  H.T.Davies,  {\it Introduction to Nonlinear Differential and Integral Equations }(Dover, New York,1962.)
\item [ [2] ]  R.M.May, {\it Stability and Competition in Model Ecosystems} (Princeton Univ.Press., Princeton, New Jersey,1974.)
\item [ [3] ]  E.C.Pielou, {\it Mathematical Ecology } (John Wiley and Sons, New York, 1977.)
\item [ [4] ]  F.Verhulst, {\it Nonlinear Differential Equations and Dynamical Systems } (Springer Verlag, Berlin, 1990.)
\item [ [5] ]  J.D.Murray, {\it Mathematical Biology } (Springer Verlag, Berlin-Heidelberg, 1993.).

\end {itemize}

\end {document}